\documentstyle[aps,prl]{revtex}

\begin{document}

\title{Anomalous fluctuations of two-dimensional Bose-Einstein condensates}

\author{Hongwei Xiong$^{1}$, Shujuan Liu$^{1}$, and Guoxiang Huang$^{2}$}
\address{$^{1}$Department of Applied Physics, Zhejiang
University of Technology, Hangzhou, 310032, China}
\address{$^{2}$Department of Physics and Key Laboratory for Optical and Magnetic
Resonance Spectroscopy, East China Normal University, Shanghai,
200062, China}

\date{\today}

\maketitle

\begin{abstract}
\it{We investigate the particle-number fluctuations due to the
collective excitations created in a two-dimensional (2D) and a
quasi-2D Bose-Einstein condensates (BECs) at low temperature. We
find that the fluctuations display an anomalous behavior, i. e.
for the 2D BEC they are proportional to $N^2$, where $N$ is the
total number of particles. For the quasi-2D BEC, the
particle-number fluctuations are proportional not only to $N^2$
but also to the square root of the trapping frequency in the
strongly-confined direction.}

\vspace{5mm}

PACS number(s): 03.75.Fi, 05.30.Jp, 67.40.Db\\
~~~E-mail:~hongweixiong@hotmail.com\\
\end{abstract}

\vspace{10mm}

The remarkable experimental realization of Bose-Einstein condensates (BECs)
has stimulated intensive theoretical and experimental studies on weakly
interacting Bose gases \cite{NATURE}. Much attention has been paid to the
investigations of the thermodynamic properties especially the particle
number fluctuations of three-dimensional (3D) interacting Bose gases \cite
{GIO,IDZ,ILLU,MEI,KOC,JAK}. Recently, quasi-1D and quasi-2D BECs have been
realized \cite{EXPLOW,LOW2}, which with no doubt provides many new
opportunities to explore the fascinating quantum statistical property of
macroscopic quantum systems in low dimensions. For a 2D Bose gas, the
particle number fluctuations due to the exchange of atoms between condensate
and thermal atoms have been addressed recently in \cite{XIONG2D,BHA}.

It is well known that for the temperature far below the critical
temperature, the collective excitations play a dominant role for the
fluctuations of BEC. The fluctuations of condensate due to
collective excitations in 3D Bose gases have been discussed in \cite
{GIO,MEI,XIONG1}. The scaling behavior of a 2D interacting condensate confined in
a box was investigated in \cite{MEI}. In the present paper, we shall
discuss the fluctuations originated from collective excitations for
2D and quasi-2D interacting condensates confined in a magnetic trap.

For the Bose-condensed gas confined in a magnetic trap, the total number of
particles $N$ in the system is conserved and hence a canonical (or
microcanonical) ensemble should be used. Within the canonical ensemble a
fairly general method has been developed recently \cite{XIONG1,XIONG2} for
studying the thermodynamic properties of the interacting Bose-condensed
gases based on the calculation of probability distribution functions. The
purpose of this work is to investigate the particle number fluctuations of a
2D BEC due to the collective excitations within a canonical ensemble.

For this aim we first extend our previous theory developed in \cite
{XIONG1,XIONG2} on particle number fluctuations in BECs by including an
effect of the quantum depletion. For a Bose gas confined in a trap, based on
Bogoliubov theory \cite{BOG} one can obtain that, at low temperature, the
total number of particles out of the condensate due to the collective
excitations is given by

\begin{equation}
\left\langle N_{T}\right\rangle =N-\left\langle N_{0}\right\rangle
=\sum_{nl\neq 0}\left\langle N_{nl}\right\rangle =\sum_{nl\neq 0}\left[
\left( \int u_{nl}^{2}dV+\int v_{nl}^{2}dV\right) f_{nl}+\int
v_{nl}^{2}dV\right] ,  \label{bogoliubov}
\end{equation}
where $\left\langle N_{T}\right\rangle $ and $\left\langle
N_{0}\right\rangle $ are respectively the number of excited atoms and atoms
in the condensate. $f_{nl}=1/\left( \exp \left( \beta \varepsilon
_{nl}\right) -1\right) $ is the average number of collective excitation $nl$
present in the system at thermal equilibrium ($\beta =1/(k_{B}T)$, $k_{B}$
is Boltzmann constant and $T$ is temperature) , while $\varepsilon
_{nl}=\hbar \omega _{nl}$ is the energy of the collective mode characterized
by the quantum numbers $n$ and $l$. In the above equation, the term $\int
v_{nl}^{2}dV$ represents the effect of the quantum depletion which does not
vanish even at $T=0$. The quantities $u_{nl}$ and $v_{nl}$ are determined by
the following coupled equations:
\begin{eqnarray}
&&\left( -\frac{\hbar ^{2}}{2m}\bigtriangledown ^{2}+V_{ext}\left( {\bf r}%
\right) -\mu +2gn\left( {\bf r}\right) \right) u_{nl}+gn_{0}\left( {\bf r}%
\right) v_{nl}=\varepsilon _{nl}u_{nl},  \label{uvenergy1} \\
&&\left( -\frac{\hbar ^{2}}{2m}\bigtriangledown ^{2}+V_{ext}\left( {\bf r}%
\right) -\mu +2gn\left( {\bf r}\right) \right) v_{nl}+gn_{0}\left( {\bf r}%
\right) u_{nl}=-\varepsilon _{nl}v_{nl},  \label{uvenergy2}
\end{eqnarray}
where $V_{ext}\left( {\bf r}\right) $ is the trapping potential confining
the Bose gas, $\mu $ and $g$ are respectively the chemical potential of the
system and inter-atomic interaction constant. $n\left( {\bf r}\right) $ and $%
n_{0}\left( {\bf r}\right) $ are the density distribution of the Bose gas
and the condensate, respectively. From Eq. (\ref{bogoliubov}) we see that $%
\left\langle N_{nl}\right\rangle $ can be taken as the average number of the
excited atoms related to the collective mode $nl$, while $\left\langle
N_{nl}^{B}\right\rangle =f_{nl}$ is the average number of the collective
excitation $nl$. From the form of Eq. (\ref{bogoliubov}), we see that the
ratio between $\left\langle N_{nl}\right\rangle $ and $\left\langle
N_{nl}^{B}\right\rangle $ is given by

\begin{equation}
\eta _{nl}=\frac{\left\langle N_{nl}\right\rangle }{\left\langle
N_{nl}^{B}\right\rangle }=\frac{\left( \int u_{nl}^{2}dV+\int
v_{nl}^{2}dV\right) f_{nl}+\int v_{nl}^{2}dV}{f_{nl}}.  \label{ratio}
\end{equation}

Within the canonical ensemble, the partition function of the
system with $N$ atoms takes the form

\begin{equation}
Z\left[ N\right] =\sum^{\prime }\exp \left[ -\beta \left( N_{{\bf 0}%
}\varepsilon _{0}+\sum_{nl\neq 0}N_{nl}^{B}\varepsilon _{nl}\right) \right] .
\label{par}
\end{equation}
In the above equation, we have omitted the interaction between collective
excitations. The prime in the summation represents the condition that the
total number of atoms in the system should be conserved within the canonical
ensemble, i. e., $\sum_{nl\neq 0}N_{nl}=\sum_{nl\neq 0}\eta
_{nl}N_{nl}^{B}=N-N_{{\bf 0}}$. One should note that in Eq. (\ref{par}), $%
N_{nl}^{B}$ is the occupation number of the collective excitation $nl$ and
should be integer number. For the convenience of calculations, by separating
out the ground-state $nl=0$ from the state $nl\neq 0$, we have

\begin{equation}
Z\left[ N\right] =\sum_{N_{{\bf 0}}=0}^{N}\left\{ \exp \left[ -\beta N_{{\bf %
0}}\varepsilon _{0}\right] Z_{0}\left( N_{T}\right) \right\} ,  \label{par2}
\end{equation}
where $Z_{0}\left( N_{T}\right) $ stands for the partition function of a
fictitious system comprising $N_{T}=N-N_{{\bf 0}}$ excited atoms which
takes the form:

\begin{equation}
Z_{0}\left( N_{T}\right) =\sum_{\Sigma _{nl\neq 0}\eta
_{nl}N_{nl}^{B}=N_{T}}\exp \left[ -\beta N_{nl}^{B}\varepsilon _{nl}\right] .
\label{par-fic}
\end{equation}
The free energy of the fictitious system is $A_{0}\left( N_{T}\right)
=-k_{B}T\ln Z_{0}\left( N_{T}\right) $.

By using the developed saddle-point method proposed in \cite{XIONG1,XIONG2},
we introduce a generating function $G_{0}\left( T,z\right) $ which is given
by

\begin{equation}
G_{0}\left( T,z\right) =\sum_{N_{T}=0}^{\infty }z^{N_{T}}Z_{0}\left(
N_{T}\right) .  \label{generating}
\end{equation}
Due to the fact that there is a confinement condition $\sum_{nl\neq 0}\eta
_{nl}N_{nl}^{B}=N_{T}$ for $N_{T}$ in the above equation, we have

\begin{equation}
G_{0}\left( T,z\right) =\Pi _{nl\neq 0}\left\{ \sum_{N_{nl}^{B}=0}^{\infty
}z^{\eta _{nl}N_{nl}^{B}}\exp \left[ -\beta N_{nl}^{B}\varepsilon
_{nl}\right] \right\} =\Pi _{nl\neq 0}\frac{1}{1-z^{\eta _{nl}}\exp \left[
-\beta \varepsilon _{nl}\right] }.  \label{generate2}
\end{equation}
$Z_{0}\left( N_{T}\right) $ can be obtained by noting that it is the
coefficient of $z^{N_{T}}$ in the expansion of $G_{0}\left( T,z\right) $.
Thus, we have

\begin{equation}
Z_{0}\left( N_{T}\right) =\frac{1}{2\pi i}\oint dz\frac{G_{0}\left(
T,z\right) }{z^{N_{T}+1}}.  \label{power}
\end{equation}

\vspace{1pt}Similar to the case of the developed saddle-point method for the
ideal Bose gas \cite{XIONG1,XIONG2}, it is easy to get the following useful
relations:

\begin{equation}
N_{T}=\sum_{nl\neq 0}\frac{\eta _{nl}}{\exp \left[ \beta \varepsilon
_{nl}\right] z_{0}^{-\eta _{nl}}-1},  \label{relation}
\end{equation}
and

\begin{equation}
-\beta \frac{\partial }{\partial N_{{\bf 0}}}A_{0}\left( N_{T}\right) =\ln
z_{0},  \label{realtion}
\end{equation}
where $z_{0}$ is the well-known saddle point.

By using the above relations, the probability distribution of the system
with $N_{0}$ atoms in the condensate can be obtained based on the method
developed in Refs.\cite{XIONG1,XIONG2}. The normalized probability
distribution function is given by \cite{XIONG1,XIONG2}

\begin{equation}
G_{n}\left( N,N_{0}\right) =A_{n}\exp \left[ \int_{N_{0}^{p}}^{N_{0}}\alpha
\left( N,N_{0}\right) dN_{0}\right] ,  \label{distribution}
\end{equation}
where $A_{n}$ is a normalized constant, $N_{0}^{p}$ is the most probable
value of the atomic number in the condensate. If taking the ground-state
energy of the system as the zero point of energy, $\alpha \left(
N,N_{0}\right) $ is determined by

\begin{equation}
N_{0}^{p}-N_{0}=\sum_{nl\neq 0}\left[ \frac{\eta _{nl}}{\exp \left[ \beta
\varepsilon _{nl}\right] \exp \left[ -\alpha \left( N,N_{0}\right) \right] -1%
}-\frac{\eta _{nl}}{\exp \left[ \beta \varepsilon _{nl}\right] -1}\right] .
\label{alpha0}
\end{equation}
For temperature below the critical temperature of BEC and for $\beta \hbar
\omega _{\perp }<<1$ (this condition is satisfied in the present-day
experiments of BECs), the above equation can be approximated to be

\begin{equation}
N_{0}^{p}-N_{0}=\sum_{nl\neq 0}\left[ \frac{1}{\exp \left[ \beta \varepsilon
_{nl}/\eta _{nl}\right] \exp \left[ -\alpha \left( N,N_{0}\right) /\eta
_{nl}\right] -1}-\frac{1}{\exp \left[ \beta \varepsilon _{nl}/\eta
_{nl}\right] -1}\right] .  \label{alpha}
\end{equation}

In fact, one can get the result given by Eq. (\ref{alpha})
through a simple physical picture. Taking the ground-state energy as the
zero point of energy, the total energy of the system is then
\begin{equation}
E=\sum_{nl}N_{nl}^{B}\varepsilon _{nl}=\sum_{nl}N_{nl}\varepsilon
_{nl}^{eff},  \label{energy}
\end{equation}
where $\varepsilon _{nl}^{eff}$ can be regarded as the effective energy
level of a single particle characterized by the quantum numbers $n$ and $l$.
From Eqs. (\ref{ratio}) and (\ref{energy}), $\varepsilon _{nl}^{eff}$ takes
the form

\begin{equation}
\varepsilon _{nl}^{eff}=\frac{\varepsilon _{nl}}{\int u_{nl}^{2}dV+\int
v_{nl}^{2}dV+\int v_{nl}^{2}dV/f_{nl}}.  \label{effective energy}
\end{equation}
Thus at low temperature the system can be regarded as a fictitious
non-interacting Bose system with the effective energy level given by $%
\varepsilon _{nl}^{eff} $. The canonical partition function of the system
reads

\begin{equation}
Z_{cp}=\sum^{\prime }\exp \left[ -\beta \sum_{nl}N_{nl}\varepsilon
_{nl}^{eff}\right] ,  \label{partition}
\end{equation}
where the prime in the summation represents the condition $\Sigma _{nl\neq
0}N_{nl}=N_{T}$. It is easy to confirm that one can get the result given by
Eq. (\ref{alpha}) from this canonical partition function.\vspace{1pt}

From Eq. (\ref{alpha}), after a straightforward calculation one obtains $%
G_{n}\left( N,N_{0}\right) $

\begin{equation}
G_{n}\left( N,N_{0}\right) =A_{n}\exp \left[ -\frac{\left(
N_{0}-N_{0}^{p}\right) ^{2}}{2\Xi }\right] ,  \label{prabability1}
\end{equation}
where

\begin{eqnarray}
& & \Xi =\sum_{nl\neq 0} \left[ \left( \int u_{nl}^{2}dV+\int
v_{nl}^{2}dV\right)^2 \left( \frac{k_{B}T}{\varepsilon_{nl}} \right)^2
\right.  \nonumber \\
& & \hspace{0.8cm}+\left. 2\left( \int u_{nl}^{2}dV+\int v_{nl}^{2}dV\right)
\int v_{nl}^{2}dV\left( \frac{k_{B}T}{\varepsilon _{nl}}\right) +\left( \int
v_{nl}^{2}dV\right)^{2} \right] .  \label{xi}
\end{eqnarray}
Below the critical temperature, $N_{0}^{p}>>1$, the fluctuations of the
condensate contributed from the collective excitations are given by

\begin{equation}
\left\langle \delta ^{2}N_{0}\right\rangle =\left\langle
N_{0}^{2}\right\rangle -\left\langle N_{0}\right\rangle ^{2}=\Xi .
\label{fluc}
\end{equation}

Generally speaking, Eqs. (\ref{xi}) and (\ref{fluc}) can be used to
investigate the fluctuations originated from the collective excitations in
any dimension. Now we specify the case of 2D. For Eqs. (\ref{uvenergy1}) and
(3), $u_{nl}$ and $v_{nl}$ are given by \cite{GRIFFIN}

\begin{equation}
u_{nl}\approx -v_{nl}\approx i\sqrt{\frac{gn_{0}\left( {\bf r}\right) }{%
2\varepsilon _{nl}}}\chi _{nl}.  \label{uv}
\end{equation}
For a Bose gas confined in a harmonic potential $V_{ext}({\bf r}%
)=m\omega_{\perp}^2 (x^2+y^2)/2$, $\chi _{nl}$ and $\varepsilon _{nl}(=\hbar
\omega _{nl})$ are determined by the eigen equation

\begin{equation}
-\frac{\omega _{\perp }^{2}}{2}\nabla \cdot \left[ \left( R_{\perp
}^{2}-r_{\perp }^{2}\right) \nabla \chi _{nl}\right] =\omega _{nl}^{2}\chi
_{nl},  \label{xixi}
\end{equation}
where $R_{\perp }$ is the radius of the condensate. After a straightforward
calculation one obtains the excitation frequency $\omega _{nl}=\omega
_{\perp }\sqrt{2n^{2}+2n|l|+2n+|l|}$, and

\begin{equation}
\chi _{nl}=\frac{A_{nl}}{R_{\perp }}e^{-il\phi }H_{nl}\left( \frac{r_{\perp }%
}{R_{\perp }}\right) ,  \label{xixi1}
\end{equation}
where $A_{nl}$ is a normalized constant determined by $\int \left| \chi
_{nl}\right| ^{2}dV=1$. In the above equation. $H_{nl}\left( x\right) $
takes the form

\begin{equation}
H_{nl}\left( x\right) =x^{|l|}\sum_{j=0}^{n}b_{j}x^{2j},
\end{equation}
where $b_{0}=1$ and the coefficients $b_{j}$ satisfy the recurrence relation
$b_{j+1}/b_{j}=\left( 4j^{2}+4j+4j|l|-4n^{2}-4n-4n|l|\right) /\left(
4j^{2}+4j|l|+8j+4|l|+4\right) $.

Substituting the above results into Eqs. (\ref{xi}) and (\ref{fluc}), we
obtain the fluctuations of particle-number in the condensate for the 2D Bose
gas:

\begin{equation}
\left\langle \delta ^{2}N_{0}\right\rangle =\frac{mg}{\hbar ^{2}}\left[
\frac{N^{2}t^{2}\left( 1-t^{2}\right) }{\pi \zeta \left( 2\right) }\gamma
_{1}+\frac{N^{3/2}t\left( 1-t^{2}\right) }{\pi \left( \zeta \left( 2\right)
\right) ^{1/2}}\gamma _{2}+\frac{N\left( 1-t^{2}\right) }{4\pi }\gamma
_{3}\right] ,  \label{analyticalfluc}
\end{equation}
where $t=T/T_{c}^{0}$ with $T_{c}^{0}=\left( N/\zeta \left( 2\right) \right)
^{1/2}\hbar \omega _{\perp }/k_{B}$ the critical temperature corresponding
an ideal Bose gas in 2D. The coefficients $\gamma _{1}$, $\gamma _{2}$ and $%
\gamma _{3}$ are given by

\[
{\ \gamma _{1} =\sum_{nl\neq 0}\frac{\beta _{nl}^{2}}{\left(
2n^{2}+2n|l|+2n+|l|\right) ^{2}}, }
\]

\[
{\ \gamma _{2} =\sum_{nl\neq 0}\frac{\beta _{nl}^{2}}{\left(
2n^{2}+2n|l|+2n+|l|\right) ^{3/2}}, }
\]

\begin{equation}
\gamma _{3}=\sum_{nl\neq 0}\frac{\beta _{nl}^{2}}{\left(
2n^{2}+2n|l|+2n+|l|\right) }.  \label{gamma}
\end{equation}
with

\begin{equation}
\beta _{nl}=\frac{\int_{0}^{1}\left( 1-x^{2}\right) x\left( H_{nl}\right)
^{2}dx}{\int_{0}^{1}x\left( H_{nl}\right) ^{2}dx}.  \label{beta}
\end{equation}
By a numerical calculation we obtain $\gamma _{1}=0.87$, $\gamma _{2}=1.43$,
$\gamma _{3}=4.37$.

From Eq. (\ref{analyticalfluc}) we have the following two conclusions: (i) $%
\left\langle \delta ^{2}N_{0}\right\rangle $ is proportional to the
inter-atomic interaction constant $g$. Therefore, for an ideal Bose gas
there is no contribution to the condensate fluctuations due to the
collective excitations. This is physically reasonable because the collective
excitations, which are dominant at low temperature, originate from the
interaction between atoms. (ii) The leading term for the fluctuations of the
particle-number in the condensate are proportional to $N^{2}$. This
anomalous behaviour comes from the low-dimensional property of the system.
Noting that in the case of 3D Bose-condensed gas the fluctuations of the
condensate due to the collective excitations are proportional to $N^{4/3}$%
\cite{GIO,XIONG1}. Thus the lower the dimension of the system is, the larger
the condensate fluctuations are. Thus at low temperature a 2D Bose gas
confined in a harmonic trapping potential, which has been realized recently
by G\"{o}rlitz et al\cite{EXPLOW}, is an ideal system for the observation of
anomalous behavior of the fluctuations of particle-number in condensates.

In real experiments \cite{EXPLOW,LOW2}, the Bose gases are confined in a
quai-2D harmonic trap. For a quasi-2D Bose gas, the coupling constant is
given by $g\approx 2\sqrt{2\pi }\hbar ^{2}a_{s}/(ml_{z})$ \cite{PETROV},
which is fixed by a $s$-wave scattering length $a_{s}$ and the oscillator
length $l_{z}=\left( \hbar /m\omega _{z}\right) ^{1/2}$ in the $z$%
-direction, where $\omega _{z}$ is the trap frequency in the $z$-direction.
In this case, we have

\begin{equation}  \label{quasi}
\left\langle \delta ^{2}N_{0}\right\rangle =\frac{2\sqrt{2\pi }a_{s}}{l_{z}}%
\left[ \frac{N^{2}t^{2}\left( 1-t^{2}\right) }{\pi \zeta \left( 2\right) }%
\gamma _{1}+\frac{N^{3/2}t\left( 1-t^{2}\right) }{\pi \left( \zeta \left(
2\right) \right) ^{1/2}}\gamma _{2}+\frac{N\left( 1-t^{2}\right) }{4\pi }%
\gamma _{3}\right].
\end{equation}
From the above equation we see that, on the one hand $\left\langle \delta
^{2}N_{0}\right\rangle$ is proportional to $N^2$ (for large $N$), and on the
other hand it is also proportional to $\omega _{z}^{1/2}$ because the factor
$1/l_z$ appearing in (\ref{quasi}). This shows that the trapping frequency
in the $z$-direction plays an important role for the particle number
fluctuations due to the collective excitations. Thus one can control the
particle number fluctuations by adjusting the trapping frequency in the $z$%
-direction. Compared with the contribution due to the thermal atoms (see Eq.
(39) in \cite{XIONG2D}), the particle number fluctuations due to the
collective excitations are strongly dependent on the trapping frequency in
the $z$-direction. From the relation between $\left\langle \delta
^{2}N_{0}\right\rangle $ and $\omega _{z}$, we see that the confinement of
the Bose-condensed gas has the effect of increasing the fluctuations due to
the collective excitations. This result is consistent with the role of the
dimensionality in the fluctuations of the condensate.

In conclusion, we have studied the particle-number fluctuations of a
condensed 2D Bose gas confined in a harmonic trapping potential by using the
probability distribution obtained through a modified saddle-point method. We
have found that the condensate fluctuations are proportional to the
inter-atomic interaction constant $g$ and the square of total
particle-number of the system. This anomalous behavior of the fluctuations
makes it very promising to experimentally observe the effect of the
particle-number fluctuations. The theoretical method provided here is quite
general and can be applied to investigate the particle-number fluctuations
of quasi-1D Bose-condensed gases.

\vspace{5mm}

This work was supported by Natural Science Foundation of China under grant
Nos. 10205011 and 10274021.

\end{document}